\documentclass[aps,nofootinbib,showpacs,preprint]{revtex4-1}
\usepackage[CJKbookmarks, pdftex, bookmarksnumbered, bookmarksopen, colorlinks, citecolor=blue, linkcolor=blue]{hyperref}
\usepackage[english]{babel}
\usepackage{amstext,amsmath,amssymb,amsfonts,bbm}
\usepackage[latin1]{inputenc}
\usepackage{graphicx}
\usepackage{epstopdf}
\usepackage{amsthm}
\usepackage{tocvsec2}
\usepackage{enumerate}
\usepackage{subfigure}
\usepackage{bm}
\usepackage{color}
\usepackage{tikz}

\usepackage[squaren,Gray]{SIunits}

\newcommand{\va}{\scriptscriptstyle}
\newcommand{\van}{\scriptstyle}

\newcommand{\be}{\nopagebreak[3]\begin{equation}}
\newcommand{\ee}{\end{equation}}

\newcommand{\bee}{\nopagebreak[3]\begin{equation*}}
\newcommand{\eee}{\end{equation*}}

\newcommand{\ba}{\nopagebreak[3]\begin{eqnarray}}
\newcommand{\ea}{\end{eqnarray}}
\DeclareFontFamily{U}{rsfs}{}         
\DeclareFontShape{U}{rsfs}{m}{n}{<5> rsfs5 <6><7> rsfs7          %
  <8><9><10><10.95><12><14.4><17.28><20.74><24.88> rsfs10}{}     %
\DeclareMathAlphabet{\mathfs}{U}{rsfs}{m}{n}                     %
\newcommand{\mfs}[1]{\mathfs {#1}}                               %
\newcommand{\n}{{\nonumber}}

\newcommand{\sO}{{\mfs O}}

 \usepackage{braket}
\newcommand{\N}{\mathbb{N}}

\newcommand{\Z}{\mathbb{Z}}
\begin{document}

\title{Hawking's information puzzle: a solution realized in loop quantum cosmology}

\author{Lautaro Amadei}
\affiliation{{Aix Marseille Univ, Universit\'e de Toulon, CNRS, CPT, Marseille, France}}

\author{Alejandro Perez}
\affiliation{{Aix Marseille Univ, Universit\'e de Toulon, CNRS, CPT, Marseille, France}}

\date{\today}

\begin{abstract}
In approaches to quantum gravity, where smooth spacetime is an emergent approximation of a discrete 
Planckian fundamental structure, any standard effective field theoretical description will 
miss part of the degrees of freedom and thus break unitarity. Here we show that these expectations can be made precise in loop quantum cosmology. Concretely, even when loop quantum cosmology is unitary at the fundamental level, when microscopic degrees of freedom, irrelevant to low-energy cosmological observers, are suitably ignored, pure states in the effective description evolve into mixed states due to decoherence with the Planckian microscopic structure. When extrapolated to black hole formation and evaporation, this concrete example provides a key physical insight for a natural resolution of Hawking's information paradox.

\end{abstract}

\maketitle

\definecolor{mycolor}{rgb}{0.122, 0.435, 0.698}

General relativity combined with quantum field theory in a regime where both are expected to be good approximations imply that large isolated black holes behave like thermodynamical systems in equilibrium. They are objects close to equilibrium at the Hawking temperature that lose energy extremely slowly via Hawking radiation.  When perturbed they come back to equilibrium to a new state and the process satisfies the first law of thermodynamics with an entropy equal to $\frac14$ of the area $A$ of the black hole horizon in Planck units. Under such perturbations (which in particular can be associated also to their slow evaporation) the total entropy of the universe can only increase. Namely, \be  
\delta S=\delta S_{\rm matter}+\frac{\delta A}{4}\ge 0, 
\ee
i.e., the generalized second law holds,  
where $\delta S_{\rm matter}$ represent the entropy outside the black hole (this law can be proved to be valid under suitable simplifying assumptions \cite{Wall:2011hj} and it is believed to hold in general).

The quasi equilibrium phase of slow evaporation (which is extremely long lasting for macroscopic black holes) is only an intermediate situation before complete evaporation. This intermediate phase is predicted by general relativity as the result of gravitation collapse. The irreversibility captured by the previous equation can be associated to  the special nature of the initial conditions leading to the collapse.
 
 As emphasized by Penrose,  among others, the arrow of time comes from the special nature of initial conditions (low curvature and initially dilute matter distribution). As times passes the system evolves to form stars that eventually collapse to form black holes.  Entropy increases because, for an observer with limited capabilities in resolving points of the phase space (coarse graining) the probability distribution defining the system seems to occupy larger and larger portions of the available phase space as they are made `available' via evolution (e.g., high densities become possible via gravitational clumping, higher temperatures are possible, etc). 
 
 The perspective we want to stress here is that the story continues to be exactly the same after black holes form, but now a huge and pristine new portion of the phase space opens by the gravitational collapse: the internal singularity of the classical description beyond the event horizon. Like the lighter setting a newspaper on fire and degrading (but not destroying) the information written in it into correlations between its microscopic constituents inaccessible to the reader, the singularity brings the system in contact with the quantum gravity scale. The gravitational collapse `ignites' interactions with the Planckian regime inside the black hole horizon, and that must be (as in the burning paper) the key point for resolving the puzzle\cite{Unruh:2017uaw}  of information in black hole evaporation. This perspective was advocated in \cite{Perez:2014xca, Perez:2017cmj}. Its realization in analog models has been explored recently \cite{Liberati:2019fse}.  

It is presently hard to prove that such scenario is viable in a quantum theory of gravity simply because there is no such theoretical framework that is developed enough for tackling BH formation and evaporation in detail. However, the application of loop quantum gravity to quantum cosmology leads to a model with similar features, and where evolution across the classical singularity is well defined \cite{Bojowald:2001xe}.  In this letter we show that the scenario is precisely realized in  loop quantum cosmology (LQC).

 More precisely, we will see that even though the quantum evolution of the universe is fundamentally unitary across the big-bang, low energy observers---unable to probe the geometry all the way down to the Planck scale---would observe an apparent deviation from unitarity due to the development of correlations with the degrees of freedom hidden to their coarse grained probing capabilities. For suitable semiclassical states, we will argue, such decoherence is negligible during evolution while curvature remains low. However, decoherence is unavoidable when evolving across the big-bang because all states (no matter how low curvature is at an initial time) would go through a phase of high curvature there.  Thus, as in the related BH evaporation context,  and when hidden Planckian degrees of freedom are ignored, pure states seem to evolve into mixed states.    

In order to illustrate our point we will use the framework of unimodular gravity. The main reason for this  is that (in the cosmological FLRW context) it completely resolves the problem of time \cite{Unruh:1988in}. More precisely, the theory comes with a preferred time evolution, and thus, it is described by a Schroedinger-like equation where states of the universe are evolved by a unitary operator. Therefore, unlike the general situation in full quantum gravity, the notion of unitarity is unambiguously defined in unimodular quantum cosmology. This is the main reason why it provides the perfect framework for the discussion of the central point in this work.

When specializing to (spatially flat)  homogeneous and isotropic cosmologies and using standard LQC  variables $\nu$ (volume of the fiducial cell times $\ell_p^{-2}$) and its conjugate momentum $b$ \cite{Ashtekar:2011ni,Chiou:2010ne}, and using the dimensionless time variable $t\equiv 4$-volume elapsed by a fiducial cell in Planck units, the action of pure unimodular gravity becomes
\begin{equation}\label{accc}
S_0=  \int
\frac{1}{2} \left[
\dot{b} \nu + \frac{3 b^2 |\nu|}{2 \gamma} N   
-{\lambda} \left(N |\nu| - \frac{2}{\gamma}\right)
\right] dt.
\end{equation}
The unimodular condition follows from varying the action with respect to the Lagrange multipier $\lambda$, and fixes the lapse to $N=2/(\gamma|\nu|)$. One can show that the Hamiltonian equals the cosmological constant $\Lambda$ 
\be\label{mavi}
H=\Lambda=\frac{3}{\gamma^2}  b^{2}.
\ee

In the representation of LQC the is no $b$ operator but only the operators corresponding to finite $\nu$ translations exist \cite{Ashtekar:2006wn, Ashtekar:2011ni}; from here on referred to as shift operators \be\label{shifty} {\exp(i 2 k b)} \Psi(\nu )=\Psi(\nu-4k).\ee 
For $k=q\sqrt{\Delta} \ell_p$ and $q\in \Z$, states that diagonalize the previous shift operators, denoted $\ket{b_0; \Gamma^{\epsilon}_\Delta}$, are labelled by a real value $b_0$ and by a graph $\Gamma^{\epsilon}_\Delta$. The graph is a 1d lattice of points in the real line of the form $\nu=4n \sqrt{\Delta}\ell_p +\epsilon$ with $\epsilon\in [0,4 \sqrt{\Delta}\ell_p)$ and $n\in \N$. The corresponding  wave function is given by $\Psi_{b_0}(\nu)\equiv \braket{\nu|b_0; \Gamma^{\epsilon}_\Delta}=\exp{(-i\frac{b_0 \nu}{2})} \delta_{\Gamma^{\epsilon}_\Delta}$ where the symbol $\delta_{\Gamma^{\epsilon}_\Delta}$ means that the wavefunction vanishes when  $\nu\notin \Gamma_\Delta^\epsilon$. It follows from (\ref{shifty}) that
\be\label{eige}  {\exp(i 2 k b)} \ket{b_0; \Gamma^{\epsilon}_\Delta}=
\exp{(i2 k b_0)} \ket{b_0; \Gamma^{\epsilon}_\Delta}.\ee
 The states $\ket{b; \Gamma^{\epsilon}_\Delta}$ are eigenstates of the shift operators that preserve the lattice $\Gamma^{\epsilon}_\Delta$. Notice, that the eigenvalues are independent of the parameter $\epsilon$. i.e. they are infinitely degenerate (a key point in what follows). 

The Hamiltonian in unimodular LQC is regularized by using shift operators with $k\equiv \sqrt{\Delta} \ell_p$, where $\sqrt{\Delta}$ is the area gap in Planck units \cite{Ashtekar:2011ni}.  We have 
\begin{equation}\label{eq89b}
\Lambda_{\Delta} \equiv  \frac{3}{\gamma^2 {\Delta} \ell_p^2 } {{\sin^2\left( {\Delta^{\frac12}} \ell_p \, b\right)}},
\end{equation}
which coincides with  \eqref{mavi}  to leading (zero) order in $\ell_p^2$.
States (\ref{eige}) with $k=k_\Delta\equiv\sqrt{\Delta} \ell_p$ diagonalize \eqref{eq89b}, i.e., they are eigenstates of the cosmological constant  \be \label{edeg}
{\widehat{\Lambda}}_{\Delta} \ket{b_0; \Gamma^{\epsilon}_\Delta}=\Lambda_\Delta(b_0)\ket{b_0; \Gamma^{\epsilon}_\Delta},\ee with eigenvalues
\be\label{energy}
\Lambda_\Delta(b_0)=3\frac{{{\sin^2\left( \sqrt{\Delta} \ell_p \, b_0\right)}}}{\gamma^2{\Delta} \ell_p^2 } .
\ee
Notice that the energy (or cosmological constant) eigenvalues do not depend on $\epsilon\in[0,4\sqrt{\Delta}\ell_p)$. Thus, the energy levels are infinitely degenerate. This is not something peculiar of our model but a general property of the representation of the canonical commutation relations used in LQC. In the pure gravity case, $\Lambda$ is positive definite and bounded from above by the maximum value $\Lambda_{\rm max}=(\gamma^{2}\ell_{p}^{2}\Delta )^{-1}$. Negative $\Lambda$ solutions are possible when matter is added \cite{us}. 

It is customary in the LQC literature to restrict to a fixed value of $\epsilon$ in concrete cosmological models, as the dynamical evolution does not mix different $\epsilon$ sectors. The terminology {\em `superselected sectors'} is used in a loose way in discussions. However, these sectors are not superselected in the strict sense of the term because they are not preserved by the action of all possible observables in the model, i.e. there are non trivial Dirac observables mapping states from one sector to another\footnote{This point was independently communicated to us in the context of Dirac observables for isotropic LQC with a free matter scalar field \cite{madha}.}. The explicit construction of such observables might be very involved in general (as it is the usual case with Dirac observables); nevertheless, it is possible to exhibit them directly at least in one simple situation: the pure gravity case. In that case the shift operators (\ref{shifty}) with shift parameter $\delta$ commute with the pure gravity Hamiltonian (the Hamiltonian constraint if we were in standard LQC) and map the $\epsilon$ sector to the $\epsilon-4\delta$ sector. The analogous Dirac observables in a generic matter model can be formally described with techniques of the type used for the definition of evolving constants of motion \cite{PhysRevD.43.442}. No matter how complicated this might be in practise, the  previous pure gravity example is an existence proof of principle.  

Thus, the infinite degeneracy of the energy eigenvalues must be understood as showing the existence of additional quantum degrees of freedom in LQC. How can we think of this large degeneracy from the fact that we would expect only a two-fold one (contracting and expanding eigenstates)? Indeed that would have been precisely the case if we had quantized the model using the standard Schroedinger representation leading to the so-called Wheeler-DeWitt quantization. The answer is to be found, we claim, in the notion of coarse graining: low energy observers (those that by definition in the toy context that quantum cosmology provides would use the Wheeler-DeWitt quantization)  must be declared to be insensitive to the huge degeneracy of energy eigenstates (cosmological constant here). All these infinitely many states for a single eigenvalue of $\Lambda$ in the LQC representation must be considered as equivalent up to the two-fold degeneracy mentioned above. In other words, coarse observers only distinguish the value of the cosmological constant (energy) and whether the universe is expanding or contracting. Such coarse graining would lead in general to decoherence as we show below.

But first let us briefly discuss the effect of coupling the previous pure-gravity model to matter. 
For simplicity we use a massless scalar field as example, whose contribution to the Hamiltonian  is
$H_{\phi}={p_{\phi}^2}/({8\pi^2 \gamma^2 \ell_p^4 \nu^2})$,
and in the quantum theory becomes \cite{Ashtekar:2011ni} 
\ba\label{hf}
&& \Hat{H}_{\rm \phi} \triangleright\ket{\psi}=\frac{m p_{\phi}^2}{16 {\Delta^2} \ell_p^4} \times\\
&&\n \times  \sum_\nu  \ket{\nu}{\left(|\nu +2 \sqrt{\Delta} \ell_p |^{\frac 12}-|\nu -2 \sqrt{\Delta} \ell_p |^{\frac 12}\right)^4}\Psi(\nu,\phi),
\ea
where one of the standard \cite{Ashtekar:2011ni} loop quantitation of the inverse volume has been used. 
The time independent Schroedinger equation becomes 
\be\label{TISE}
(\Hat{\Lambda}_{0}-\frac{8\pi \ell_p^2}{V_0}\Hat H_{\rm \phi}) \triangleright  \ket{\psi}=\Lambda  \ket{\psi}. 
\ee
The momentum $p_{\phi}$ commutes with the total Hamiltonian and thus it is a constant of motion.
Therefore, for an eigenstate of $p_{\phi}$, the previous equation is equivalent to a scattering problem in $1d$ quantum mechanics with the variable $b$ playing the role of momentum of a particle in a potential that decays as $\nu^{-2}$. 

 This problem is studied further in \cite{us}. However, the main objective of our present analysis is to illustrate an idea in terms of a concrete and simple toy model. In this sense it seems easier to modify the structure suggested by the  scalar field coupling and simply replace it by an interaction where the `long distance' behaviour  of the function $F(\nu; \lambda)$ is replaced by a short range analog $F(\nu; \lambda)\propto \delta_{\nu,0}$. The qualitative properties of the scattering will be the same and the model becomes sufficiently trivial for straightforward analytic computations.

Therefore, we consider an interaction 
\begin{equation}\label{eq147}
    \Hat{\Lambda} = \Hat{\Lambda}_{0} - \mu \frac{8\pi \ell_p^2}{V_0}\Hat{H}_{\rm int},
\end{equation}
where $\mu$ is a dimensionless coupling, $\Hat{\Lambda}_{0}$ is the pure gravity Hamiltonian, and 
$\Hat{H}_{\rm int}$ is defined as
\be\label{eqq36}
\Hat{H}_{\rm int} \triangleright \ket{\psi}\equiv \sum_\nu  \left( \ell^{-4}_p \frac{V_0}{\sqrt{\Delta}} \right)\ket{\nu} \frac{\delta_{\nu,0}}{\sqrt{\Delta}} \Psi(0).
\ee
We have added by hand an interaction Hamiltonian that switches  on only when the universe evolves through the {\em would-be-singularity} at the zero volume state. The key feature of the $\Hat{H}_{\rm int}$ is that---as its more realistic relatives matter Hamiltonian (\ref{hf})---it breaks translational invariance and in that it will lead to different dynamical evolution for different $\epsilon$-sectors. 

Before and after the big-bang the state of the universe is described by the eigenstates of the pure gravity Hamiltonian (\ref{edeg}). One needs to take special care of the peculiar degeneracy of energy eigenvalues contained in the $\epsilon$-sectors.
We will consider, for simplicity, the superposition of only two lattices $\Gamma_\Delta^\epsilon$  with $\epsilon=0$ for the first one and $\epsilon=2\sqrt{\Delta} \ell_p$ for the second one. The degenerate eigenstates of the shift operators (\ref{eige}) with eigenvalues $\exp(i2kb)$ will be denoted
\ba\label{dedito}
\ket{b,1}\equiv \ket{b;\Gamma_\Delta^0},\ \ \ \ 
{\rm and} 
\ \ \ \ 
\ket{b,2}\equiv \ket{b;\Gamma_\Delta^{2\sqrt{\Delta} \ell_p}},
\ea
respectively, while we will denote by $\Gamma^1$ and $\Gamma^2$ the corresponding underlying lattices. The immediate observation is that states supported on $\Gamma^2$ (superpositions of $\ket{b,2}$) will propagate freely because they are supported on a lattice that does not contain the point $\nu=0$. On the other hand, states supported on $\Gamma^1$ (superpositions of $\ket{b,1}$)  will be affected by the interaction at the big-bang. Such asymmetry of the interaction on different $\epsilon$-sectors is not an artifact of the simplicity of the interaction Hamiltonian. This is just a consequence of the necessary breaking of the shift invariance of the matter Hamiltonian that is a generic feature of any realistic matter coupling.

Therefore, the non trivial scattering problem concerns only states on the lattice $\Gamma^1= \{\nu= 2n\sqrt{\Delta} \ell_p \ | \ n \in \mathbb{Z} \}$ that is preserved by the Hamiltonian and contains the point $\nu=0$. As in standard scattering theory we consider an in-state of the form
\begin{equation}\label{eq151}
\ket{\psi_k}  =
\ket{\nu}\begin{cases}
e^{- i \frac k 2\nu} + A(k) \, e^{i \frac{k}{2} \nu} & \text{($\nu \ge 0$)} \\
B(k)\, e^{- i \frac k2 \nu} & \text{($\nu \le 0$)},
\end{cases}.
\end{equation}
where $\nu\in \Gamma^1$,  and $A(k)$ and $B(k)$ are coefficients depending on $k$.  For suitable coefficients, such states are eigenstates of the Hamiltonian $H_0$ as well as the full Hamiltonian \eqref{eq147}.
Arbitrary solutions (wave packets) can then be constructed in terms of appropriate superpositions of these `plane-wave' states.
   
 We can compute the scattering coefficients $A(k)$ and $B(k)$ from the discrete time-independent Schrodinger equation \eqref{TISE}, which becomes the following difference equation in the $\nu$ basis:   
 \ba\label{shishi}
\n && 3\sum_\nu \frac{\Psi(\nu - 4\sqrt{\Delta} \ell_p) + \Psi(\nu + 4\sqrt{\Delta} \ell_p)  - 2\Psi(\nu) }{2 \gamma^2\Delta \ell_p^2}   \n \\ && \ \ \ \ \ \  \ =\sum_\nu \left(\frac{8\pi \mu}{{\Delta} \ell_p^2} \delta_{\nu,0} \Psi(0) - \Lambda(k)\,  \Psi(\nu)  \ket{\nu}\right),
\ea
 The solution of the previous equations is
\ba \label{eq152}
    A(k) = \n \frac{- i \Theta(k)}{1 + i \Theta(k)}, \ \ \ \
     B(k) = \frac{1}{1 + i \Theta(k)}.\ea
where    
\ba      \label{61-s}
     \Theta(k) \equiv  \frac{16\pi \gamma^2} {3} \frac{\mu}{\sin(2 k \sqrt{\Delta} \ell_p)}.
\ea
We consider an ${\rm in}$-state of the form (valid for early times)
\vskip-.1cm
\ba\label{42}
&& \ket{\psi_{\rm in}, t}=  \\ && \n \frac{\pi}{\sqrt{ 2\Delta} \ell_p} \int db \big( \ket{b,1} \psi(b)+\ket{b,2} \psi(b) \big) e^{-i \Lambda_{\Delta}(b) t },
\ea
where $\psi(b)$ is a wave function picked at some $b=b_0$ value and $\nu=\bar \nu$.  Notice that we are superimposing two wave packets supported on lattices $\Gamma^1$ and $\Gamma^2$ respectively.  
Let us assume that $\psi(b)$ is highly picked at some $b_0$ so that we can substitute the integration variables $b$ and $b^\prime$ by $b_0$ and have a finite dimensional representation of the reduced density matrix after the scattering (this step is rather formal, it involves an approximation but it helps visualising the result).
In the relevant $4\times4$ sector we get
\vskip-.4cm\ba\label{iny}
&& {\bm \rho}_{\rm in}=\left(
\begin{array}{cccc}
 \van \frac{1}{2} &\van  0 & \van \frac{1}{2} &\van  0 \\
\van  0 &\van  0 &\van  0 &\van  0 \\
 \van \frac{1}{2} &\van  0 &\van  \frac{1}{2} &\van  0 \\
 \van 0 & \van 0 &\van  0 &\van  0 \\
\end{array}
\right), \ea
and
\ba\label{outy}
{\bm \rho}_{\rm out}=\van \frac{1}{2}\left(\begin{matrix}  {\van |B(b_0)|^2}& \van{\overline A(-b_0) B(b_0)} & {\van B(b_0)}  & \van 0 \\
\van A(-b_0) \overline{B(b_0)} &\van |A(-b_0)|^2  & \van A(-b_0) & \van 0\\
\van \overline{B(b_0)}& \van \overline {A(-b_0)}& \van 1 & \van 0 \\
 \van 0& \van 0&\van  0&\van  0 \end{matrix}\right),
\ea
respectively. Both of which are pure states.

The reduced density matrix, representing the state as seen by coarse observers (only sensitive to $b$), is defined by tracing over the discrete degree of freedom labeling the component of the state in either the $\Gamma_1$ or the $\Gamma_2$ lattices, namely \vskip-.6cm
 \be 
\bra{b}\rho^{\rm \va R}\ket{b^{\prime}}\equiv \sum_{i=1}^2\bra{b,i}\rho\ket{b^\prime,i}.\ee
From \eqref{outy} we get the reduced density matrix 
\be
{\bm \rho}^{\va \rm R}_{\rm out}=\van \frac{1}{2}\left(\begin{matrix}\van   1+ |B(b_0)|^2 & \van \overline A(-b_0) B(b_0)  \\
\van  A(-b_0) \overline{B(b_0)} &\van   |A(-b_0)|^2\end{matrix}\right),
\ee
which is now mixed, while ${\bm \rho}^{\va \rm R}_{\rm in}$ remains pure. Hence, the entanglement entropy jumps. The result (for small $\Lambda$) is
\be
\delta S=\log (2)-\frac{3 \Delta  
   }{128 \pi ^2 \gamma ^2 \mu ^2}  \Lambda \ell_p^2+\sO(\Lambda^2 \ell_p^4)\ee
The entropy jumps at the big-bang no matter how small scalar curvature the initial state has. Low energy observers (insensitive to the UV details) would conclude that pure states evolve into mixed states.

The sudden jump here is only an artifact of the ultra-local simplification of the matter contribution to the Hamiltonian in \eqref{TISE}. It is easy to see that in the more realistic situation---where for example matter is modelled by the free scalar field Hamiltonian \eqref{hf}---entanglement entropy will grow gradually as we approach the big-bang region where the matter Hamiltonian changes more and more rapidly with $\nu$ and hence the components of the quantum state on different $\epsilon$-sectors `see' different potentials. 
This effect depends on the curvature of the semiclassical state. More precisely, initially when the universe is large and for low $\Lambda$, decoherence will become important closer to the big-bang than for higher $\Lambda$ states simply because for the wave function oscillates faster for higher $b$ amplifying the differences between the matter Hamiltonian when proved with two different $\epsilon$-sectors. This is confirmed by numerical solutions of \eqref{TISE}  \cite{us}.

A different possibility that we have investigated  corresponds to the case where the system is generalized by promoting the area gap $\Delta$ to a quantum operator acting on a Hilbert space that is the tensor product of the usual LQC Hilbert space and a finite dimensional one defined by the span of a finite number of eigenstates of $\Delta$ (for concreteness we take the span of two states with eigenvalues $\Delta$ and $4\Delta$ respectively). This additional quantum number represents a UV degree of freedom entering the so-called $\bar\mu$-regularization scheme \cite{Ashtekar:2011ni} that remains  hidden to low energy observers. 

A simple calculation shows that a state of the universe without correlations with such underlying UV data will unavoidably develop correlations in the future \cite{us}. The entanglement entropy defined by tracing out the hidden degrees of freedom grows from zero with time even in the pure gravity case. This can be see from the expansion 
of the energy eigenvalues (\ref{eq89b}) in powers of $b^2$, namely
\begin{equation}\label{eq158}
\Lambda_{\Delta}(b)=  \frac{3}{\gamma^2}   b^2-\frac{1}{\gamma^2}  \Delta \ell_p^2 b^4+b^2\sO(\ell_p^4 b^4),
\end{equation}
and the fact that the second term should be interpreted as an interaction Hamiltonian between the macroscopic degree of freedom $b$ and the UV hidden one $\Delta$. In the pure gravity case decoherence can be made as small as wanted by looking at states with sufficiently low curvature (unitarity is recovered for small curvature states in Planck units). However, when matter is coupled to gravity then (once more) entanglement entropy jumps  across the big-bang independently of the initial curvature. For our ultra-local matter model \eqref{eqq36} the result (to leading order in $\Lambda$) is     
\be\label{bbb}
\delta S=\delta_0 S-\frac{3
   \Delta \log (3)}{128 \pi
   ^2 \gamma ^2 \mu ^2}\ \Lambda  \ell_p^2 +\sO(\Lambda^2\ell_p^4),
\ee
where $\delta_0 S=2\log(2)-\frac34 \log(3)$.

In conclusion, the quantization techniques of loop quantum gravity applied to cosmology yield a Hilbert space that is vastly larger that the one of the Wheeler-DeWitt representation. As a consequence, physical states labelled by macroscopic quantum numbers carry an extra degeneracy associated to microscopic or Planckian degrees of freedom. This is believed to be the case in the full theory also, where, for example, states corresponding to flat spacetime are expected to be degenerate.  The existence of such microscopic degrees of freedom are actually the one responsible for black hole entropy in this approach \cite{Perez:2017cmj, G.:2015sda}. Such {\em reservoir} of quantum numbers remains hidden in any effective QFT approach to BH evaporation and can resolve the information puzzle \cite{Perez:2014xca, Perez:2017cmj}. In this work we show that this is precisely the case for a quantum cosmology model where entanglement with the (also present) microscopic structure leads to the apparent loss of information for low energy observers in a framework where the fundamental theory is unitary.

These results extrapolated to the context of black hole formation and evaporation provide a possible resolution of the information paradox  that avoids pathological features as firewalls \cite{Almheiri:2012rt, Braunstein:2009my} or the risks of information cloning that holographic type of scenarios must deal with \cite{Marolf:2017jkr}. As decoherence in our model takes place without diffusion \cite{Unruh:2012vd}, the usual difficulties \cite{Banks:1983by} with energy conservation in the purification process are avoided along the lines of \cite{Unruh:1995gn, Unruh:2012vd}, yet in a concrete quantum gravity framework (hence without the problems faced by the QFT approach \cite{Hotta:2015yla, Wald:2019ygd}). For an alternative approach involving modifications of quantum mechanics see \cite{Modak:2014vya, Okon:2016qlh, Okon:2017pvc}. 

\bibliography{referencias}
\bibliographystyle{unsrt}

\end{document}